\documentclass[12pt,thmsa,a4paper]{article}

\usepackage{amsmath}
\usepackage{amssymb}
\usepackage{cite}
\usepackage{rotating}

\newcommand{\plabel}{\label}

\newcommand{\beqs}{\begin{equation*}}
\newcommand{\beq}{\begin{equation}}

\newcommand{\eeqs}{\end{equation*}}     
\newcommand{\eeq}{\end{equation}}

\newcommand{\beqas}{\begin{eqnarray*}}
\newcommand{\beqa}{\begin{eqnarray}}

\newcommand{\eeqas}{\end{eqnarray*}}
\newcommand{\eeqa}{\end{eqnarray}}

\newcommand{\eq}[2]{\begin{equation} #1 \plabel{#2} \end{equation}}

\newcommand{\eqa}[2]{\begin{eqnarray} #1 \plabel{#2} \end{eqnarray}}

\newcommand{\meq}[2]{\begin{multline} #1 \plabel{#2} \end{multline}}

\newcommand{\eps}{\varepsilon}
\newcommand{\al}{\alpha}
\newcommand{\be}{\beta}
\newcommand{\ga}{\gamma}
\newcommand{\de}{\delta}
\newcommand{\om}{\omega}
\newcommand{\ka}{\kappa}
\newcommand{\la}{\lambda}

\newcommand{\vphi}{\varphi}

\newcommand{\Om}{\Omega}

\newcommand{\blist}{\begin{itemize}}

\newcommand{\elist}{\end{itemize}}

\begin{document}

\begin{titlepage}
\renewcommand{\thefootnote}{\fnsymbol{footnote}}

\hfill TUW--00--16 \\

\begin{center}

%% \textbf{\Large Two Dilaton Theories in Two Dimensions}\\
{\Large\bf Two-Dilaton Theories in Two Dimensions\newline
from Dimensional Reduction}\\
\vspace{3ex}

Revised version to be published in the Annals of Physics.

  \vspace{7ex}
  D.~Grumiller\footnotemark[1],
  D.~Hofmann\footnotemark[2],
  W.\ Kummer\footnotemark[3],
  \vspace{7ex}

  {\footnotemark[1]\footnotemark[2]\footnotemark[3]\footnotesize Institut f\"ur
    Theoretische Physik \\ Technische Universit\"at Wien \\ Wiedner
    Hauptstr.  8--10, A-1040 Wien, Austria}
  \vspace{2ex}

   \footnotetext[1]{E-mail: \texttt{grumil@hep.itp.tuwien.ac.at}}
   \footnotetext[2]{E-mail: \texttt{hofmann@hep.itp.tuwien.ac.at}}
   \footnotetext[3]{E-mail: \texttt{wkummer@tph.tuwien.ac.at}}
\end{center}
\vspace{2ex}

\begin{abstract}
  Dimensional reduction of generalized gravity theories or 
 string theories generically  yields 
  dilaton fields in the lower-dimensional effective theory. 
  Thus at the level of  D=4
  theories, and cosmology  many models contain more than
  just one scalar field  (e.g.\  inflaton, Higgs,
  quintessence).  Our present work is  restricted to 
  two-dimensional gravity theories with only two
  dilatons which nevertheless allow a large class of physical 
  applications. \\ 
  The notions of
  factorizability, simplicity and conformal simplicity,
  Einstein form and Jordan form are the basis of an adequate
  classification.   We show that practically all
  physically motivated models belong either to the
  class of factorizable simple theories (e.g.\ dimensionally
  reduced gravity, bosonic string) or to factorizable
  conformally simple theories (e.g.  spherically reduced
  Scalar-Tensor theories).
  For these theories a first order formulation is 
  constructed straightforwardly. As a consequence 
  an absolute conservation law can be established. 
\end{abstract}

\\
\vfill

\end{titlepage}

\section{Introduction}

Dilaton fields have experienced an impressive comeback in recent 
years in a broad range of gravitational theories. 
Motivated by their appearance in string theories 
scalar fields play an increasingly important
role in modern physics. In the context of those theories, 
but also as a feature of any higher-dimensional theory of 
gravity, the concept of compactification has become
a standard method in many models that leads inevitably to
the occurrence of dilatons in the reduced action.
It is the aim of this paper to discuss some common properties
of two-dimensional two-dilaton theories (TDT) where at least
one dilaton field is produced by dimensional reduction.

Historically the dilaton was introduced for the first time 
by  Kaluza and Klein who proposed a five-dimensional gravity theory (KKT)
to unify general relativity with electrodynamics
\cite{kal21}.  The scalar field created by
reduction to 4 dimensions inspired Fierz \cite{fie56} and Jordan 
\cite{jor55} to invent the first Scalar-Tensor
theory (STT) in 4D where the dilaton field was interpreted as a
local gravitational coupling ``constant''.  
Already Fierz \cite{fie56} investigated the connection of this 
theory with usual general relativity by conformal transformations.  
Later work of Brans and Dicke \cite{brd61} revived the theory which in the 
following will be called Jordan-Brans-Dicke theory (JBD). 
Recently the interest in STT has increased enormously  due to the observation 
of accelerating galaxies with high redshift, indicating a positive 
cosmological constant \cite{pal97}. Again the transformation of that constant 
into a scalar field (``quintessence'') has been proposed \cite{zws99}.

A one-dilaton theory in two dimensions emerges naturally 
\cite{tih84} in connection with spherically reduced general
relativity (SRG). The reformulations of general one-dilaton theories in D=2 
as first order theories with torsion \cite{iki93} 
have led to various new insights, including e.g.\ the discovery of a 
conservation law \cite{kus92}. These results have been extended 
to the case of SRG with a massless scalar field minimally coupled in the 4D 
theory, i.e.\  the Einstein massless Klein-Gordon model (EMKG) 
\cite{kut98, grk00}.
 
To motivate the interest in TDT we briefly summarize already existing models 
that belong to this category:

\blist
\item The most obvious example is spherically reduced EMKG.  
Whenever one deals with a scalar field in ordinary general relativity and 
demands spherical symmetry one arrives at a TDT in
2D, where the 4D scalar field may be interpreted as one of the two dilatons. 
\item The polarized Gowdy model \cite{gow71} 
is based on the existence of two commuting space-like Killing fields
in a closed Einstein universe. Then toric reduction directly leads to a
TDT.  This example is of particular interest because here
both dilatons in the 2D theory are part of four-dimensional
geometry.  The polarized Gowdy model (in contrast to SRG)
allows to retain one degree of freedom of the gravity waves 
which is transferred into one of the dilatons.
\item Another example is given by KKT.  Having already one
dilaton in 4 dimensions one ends up again with a TDT in 2D
through spherical reduction.  This is not equivalent to spherically reduced
EMKG since the Kaluza dilaton in the four-dimensional theory couples
non-minimally to gravity.
\item STT and nonlinear gravity theories are mainly rooted in KKT, thus  
their connection to TDT is very similar. In addition it has been shown that 
nonlinear gravity theories are formally equivalent to STT (cf. e.g. 
\cite{mas94}).  
\elist

Finally, TDT in 2D may serve as useful toy models for TDT in 4D which may be 
necessary to describe the appearance of various scalar fields, encountered in 
cosmology (Higgs, inflaton, stringy dilaton(-s), quintessence). Up to now 
there exists no 4D STT with a single scalar field playing the role of 
{\em all} of these fields and certain dimensional reductions of e.g. a 11D 
supergravity \cite{dnp86} can yield such theories by analogy to the 
$5 \to 4 \to 2$ reduction of spherically reduced KKT. 

Thus, a dilaton field either may be produced by dimensional reduction (like in 
the effective theories for the Gowdy model or KKT) or it is introduced in the 
action as a generic scalar field (like in STT or the EMKG).

In section 2 we present the general framework of the TDT.
The notions of Einstein form and Jordan form are introduced. As examples three
significant physical applications are shown to fit into this framework.

In section 3 a useful classification scheme is invented distinguishing
models that are simple and/or factorizable. We further investigate
how conformal transformations affect these properties.

Section 4 is devoted to (conformally) simple factorizable theories. They can 
be treated in a first order form where a conservation law easily can be 
derived. Finally we examine the scaling properties of the conserved quantity. 

In the Conclusions a table summarizes the various models we consider, together
with their properties. Finally possible further applications are discussed.

\section{General framework}

The ansatz for the TDT
\eqa{
S_a &=& \int_{M_2} d^2x \sqrt{-g} \left[ V_0 (X,Y) R + V_1 (X,Y) \nabla_{\al} X
\nabla^{\al} X + V_2 (X, Y) \nabla_{\al} Y \nabla^{\al} Y \right. \nonumber \\
&& \left. + V_3 (X,Y) \nabla_{\al} X \nabla^{\al} Y + V_4 (X,Y) + V_5 (X,Y) 
f_m(S_n, \nabla_{\al} S_n, \dots) \right]
}{1}
satisfies diffeomorphism invariance in 2D and the following requirements:
\blist
\item Two scalar dilaton fields $X, Y$ should appear in the 2D action.
\item The action should be linear in the scalar curvature $R$ since 
terms with higher power in $R$ could be accommodated by modifying
the arbitrary functions in (\ref{1}) just like in theories 
with only one dilaton field \cite{mas94}. In order to have a 
nontrivial dilaton geometry the factor $V_0$ is assumed never to 
be a constant. 
\item The dilaton fields' first derivatives enter
quadratically multiplied by an arbitrary function of the
dilatons ($V_1$ and $V_2$); in general, there will be a
mixing between them ($V_3 \neq 0$).
\item In addition, there is an arbitrary function of the
dilaton fields, $V_4$, henceforth called ``potential''.
\item Finally, there are contributions from one or more ``matter 
fields'' $S_n$ which couple non-minimally to the dilatons whenever 
$V_5 \neq\mbox{const.}$
\elist

Our paper  will mainly deal with the special case $V_5=0$, 
for simplicity, although when conformal transformations
are discussed we will have to reconsider the matter
part since the coupling function $V_5$ will change in general, having important
implications for geodesics and hence for the global structure 
of the manifold.

\subsection{Standard forms}

In the following we restrict the function $V_0(X,Y)$ further, in 
accordance with our intention to examine dimensionally reduced gravity 
models. The application of conformal transformations in the 
higher-dimensional theory reveals two standard forms, which 
have the advantage that all models considered in the Introduction fit 
into one of them. Bearing in mind that at least one of the dilatons 
(here called $X$) possesses a conformal weight $\alpha\neq 0$ (being 
part of the higher-dimensional metric), we cannot reach a form where 
$V_0=const.$ through this kind of conformal transformation.

\subsubsection{Einstein form}
We call the first standard form ``Einstein form''(EF) because it contains as 
the most important representative SRG in the Einstein frame in D=4. 

It is given by
\begin{multline}
S_E = \int_{M_2} d^2x \sqrt{-g} \left[ X R + V_1^E (X,Y) \nabla_{\al} X 
\nabla^{\al} X + V_2^E (X, Y) \nabla_{\al} Y \nabla^{\al} Y \right. \\
\left. + V_3^E (X,Y) \nabla_{\al} X \nabla^{\al} Y + V_4^E (X,Y) + 
V_5^E (X,Y) f_m(S_n, \nabla_{\al} S_n, \dots) \right] \label{2}
\end{multline}
and contains important special cases listed in table \ref{tab:table1}
at the end of the paper. A trivial subclass are one-dilaton theories where 
$V_i^E = V_i^E(X)$ and $V_2^E = V_3^E = V_5^E = 0$.

\subsubsection{The Jordan form}

Motivated by spherically reduced JBD in the Jordan frame, we call the 
second standard form ``Jordan form'' (JF).  It reads
\begin{multline}
S_J = \int_{M_2} d^2x \sqrt{-g} \left[ XY R + V_1^J (X,Y) \nabla_{\al} X 
\nabla^{\al} X + V_2^J (X, Y) \nabla_{\al} Y \nabla^{\al} Y \right. \\
\left. + V_3^J (X,Y) \nabla_{\al} X \nabla^{\al} Y + V_4^J (X,Y) + 
V_5^J (X,Y) f_m(S_n, \nabla_{\al} S_n, \dots) \right] . \label{3}
\end{multline}
Indeed its most important representatives are general spherically 
reduced STT. There 
$V_1^J = Y/(2X)$, $V_2^J = -w X/Y$, $V_3^J = 2$, 
$V_4 = - 2Y + X \hat{V}(Y)$ and $V_5 = X$. In JBD $w = const.$ is called
``Dicke parameter''. $\hat{V}(Y)$ is a scalar potential (which vanishes in
JBD) and $V_5$ is chosen such that it amounts to minimal
coupling of matter fields in the Jordan frame of four-dimensional STT.

\subsection{Applications: Specific models}

In this section we consider three significant models somewhat more in detail.  
They are all constructed through dimensional reduction of
D-dimensional gravity theories by assuming the existence of
(D-2) spacelike Killing fields.  In the first case we start
from the spherically symmetric D-dimensional Einstein-Hilbert action
with one (in D dimensions minimally coupled) massless scalar field. In the
second case we apply the spherical reduction scheme to STT in D = 4 
without matter where the scalar field plays the role of one 
of the two dilatons in two dimensions. In a last example we 
reduce the pure 4D polarized Gowdy model that has cylindrical 
symmetry and therefore one gravitational degree of freedom.  

\subsubsection{Spherically reduced Einstein gravity with massless scalar field}

The D-dimensional (D$\ge$4) Einstein-Hilbert action including a massless 
scalar field $Y$ reads
\eq{
S=\int_{M_D}d^Dx\sqrt{-g}\left[R^{(D)}-\kappa\nabla_{\mu}Y\nabla^{\mu}Y\right]
.}
{D1}
In D=4 Einstein gravity the constant $\ka$ is taken to
be $16\pi G$ with Newton's constant $G$. If the D-dimensional
spacetime $M_D$ is spherically symmetric, its metric
can be written as\footnote{We use the metric
signature $(+, -, \dots)$.  The indices $\alpha,\beta,\gamma$ take the
values $0,1$ whereas the indices $\kappa,\lambda$ run from
$2$ to $D-1$. Indices $\mu,\nu$ run from $0$ to $D-1$.} 
\begin{equation}
ds^{2}=g_{\alpha\beta}dx^{\alpha}dx^{\beta}-X^2\left(r,t\right)
g_{\kappa\lambda}dx^{\ka}dx^{\la} ,
\end{equation}
where $g_{\alpha\beta}$ is the metric of a two-dimensional Lorentz 
manifold $M_2$, $g_{\ka\la}$ the metric of a (D-2)-sphere and $X$ 
the dilaton field. The scalar curvature $R^{(D)}$ of $M_D$ can be 
decomposed as (cf. e.g. \cite{tih84})
\eq{
R^{(D)}= R-\frac{(D-2)(D-3)}{X^2}\left[1+\nabla_{\alpha}X\nabla^{\alpha}X
\right]-2(D-2)\frac{\square X}{X}}
{D2}
where $R$ on the right side is the curvature of $M_2$.  To integrate out the 
isometric angular coordinates on the unit sphere $S^{D-2}$ we only have to 
substitute the scalar curvature by the above expression and the measure by 
\eq{
\sqrt{(-1)^{D+1}g_{M_D}}=\sqrt{-g}\,\sqrt{(-1)^Dg_{S^{D-2}}}\cdot
X^{D-2} .} {D2a} For later convenience we perform a
field redefinition \eq{ X\rightarrow (D-2)X^{1/(D-2)} .}
{D2b}
Up to a constant factor, the effective 2D action thus reads 
\eq{
S=\underset{M_2}{\int}d^2x\sqrt{-g}\left[XR+\frac{(D-3)}{(D-2)}
\frac{(\nabla X)^2}{X}-\frac{(D-3)}{(D-2)}X^{\frac{(D-4)}{(D-2)}}-\ka
X(\nabla Y)^2\right]} {D3}
and obviously is of the Einstein form (\ref{2}). 

\subsubsection{Spherically reduced Scalar-Tensor theories}

The four-dimensional STT action without matter is given by 
($\phi \in \mathbb{R}^+$)
\eq{
S_{STT}=\int_{M_4}d^4x\sqrt{-g}\left[\phi R-w(\phi)
\frac{\nabla_{\mu}\phi\nabla^{\mu}\phi}{\phi}+V(\phi)\right] . 
}{D13}
Here $\phi$ is the (positive) scalar field (STT field) that couples 
non-minimally to the metric. In the original KKT this non-trivial
coupling was a result from a former reduction of a five-dimensional theory. 
In STT one ``forgets''
about this fact and instead employs this coupling by hand. 
In JBD $w$ is an arbitrary constant whereas in recent 
quintessence theories a dependence $w(\phi)$ has been proposed. A 
phenomenological potential $V(\phi)$ can be used to describe various 
cosmological scenarious \cite{dan93}. Spherical reduction occurs similar to 
the case of Einstein gravity. Replacing the scalar curvature by (\ref{D2}), 
using the field redefinition (\ref{D2b}) and setting D=4 we can integrate out 
the angular coordinates on $S^2$, i.e. $\theta,\vphi$, to obtain the 2D action
\meq{
S=\int_{M_2}d^2x\sqrt{-g}\left[\phi\left(XR+\frac{(\nabla X)^2}{2X}-
\frac{1}{2}\right)+2\nabla_{\alpha }\phi \nabla^{\alpha}X \right. \\
\left. -w(\phi)X\frac{(\nabla \phi)^2}{\phi}+XV(\phi) \right] .
}{D14}
Here we have already performed a partial integration and divided by the
overall factor $4\pi$. It is now convenient to apply a conformal transformation
\eq{
g_{\mu\nu}\rightarrow g_{\mu\nu}\phi^{-1}\,\,\,,\,\,\,X\rightarrow X\phi^{-1} ,
}{D15}
together with a field-redefinition $\phi = A(Y)$, $A$ being a solution 
of the ordinary differential equation
\eq{
\frac{d \ln A}{d Y} = \left(w(A)+\frac{3}{2}\right)^{-1/2}
}{D15a}
that brings the action to the Einstein form
\eq{
S_{E}=\int_{M_2}d^2x\sqrt{-g}\left[XR+\frac{(\nabla X)^2}{2X}-\frac{1}{2}-
X(\nabla Y)^2+\frac{X}{A(Y)}V(A(Y))
\right] .}
{D16}
In this form the mixed term $\nabla_{\al}\phi\nabla^{\al}X$ 
disappears. However, in the case of interaction with matter a 
complicated nonminimal coupling to the STT field arises. In the matterless 
case the STT field is seen simply to play the role of an additional scalar 
field with proper (nonminimal) coupling to the geometric dilaton $X$ in D=2.

\subsubsection{Gowdy model}

The four-dimensional (polarized) Gowdy metric \cite{gow71}
\eqa{
ds^2&=&e^{2a(t, \theta)}\left(dt^2-d\theta^2\right)-X(t, \theta)
\left(e^{Y(t, \theta)}d\sigma^2+e^{-Y(t, \theta)} d\delta^2\right).}
{D4}
describes a 4D spacetime that has 2 commuting Killing fields spanning a flat, 
spacelike isometry submanifold $T^2\cong S^1\otimes S^1$ (locally). 
Moreover it is assumed that the whole spacetime $M_4$ is compact.
Performing the integration over the isometric 
coordinates $\sigma, \delta$ yields an effective two-dimensional action. For 
this reason we have to decompose the 4D scalar curvature $R^{(4)}$ into terms 
corresponding to $T^2$ and $M_2$, which is the complementary manifold, and 
terms produced by the embedding. This computation is done most conveniently in 
the vielbein frame\footnote{Letters from the Latin alphabet denote vielbein 
indices, while Greek ones are reserved for coordinate indices.}
$ds^2 =\eta_{ab}e^ae^b-\left(e^2\right)^2-\left(e^3\right)^2$.
Quantities associated to $T^2$ or $M_2$ shall be assigned a tilde. 
We treat $T^2$ as two independent one-dimensional spaces $S_1$. 
Thus the relation between the vielbeine is given by
\eq{
e^a = \tilde{e}^a,\,\, e^2 = \sqrt{X}e^{\frac{Y}{2}}\tilde{e}^2,\,\, 
e^3=\sqrt{X}e^{-\frac{Y}{2}}\tilde{e}^3 .
}{D6}
Demanding vanishing torsion and metric compatibility on $M_4, M_2$ and $T^2$ 
the connection 1-form on $M_4$ is obtained:
\eqa{
\omega^a_{\,\,b}=\tilde{\omega}^a_{\,\,b}&,&\,\,\,\omega^2_{\,\,3}=
\tilde{\omega}^2_{\,\,3}=0\\
\omega^2_{\,\,a}=\left(\tilde{E}_a\sqrt{X}e^{\frac{Y}{2}}\right)
\tilde{e}^2&,&\,\,\,\omega^3_{\,\,a}=\left(\tilde{E}_a\sqrt{X}e^{-\frac{Y}{2}}
\right)\tilde{e}^3.
\nonumber}
{D7}
This is sufficient to calculate the scalar curvature
\eq{
R^{(4)}= R-2\frac{\square X}{X}+\frac{1}{2}
\frac{\nabla_{\al}X\nabla^{\al}X}{X^2}-\frac{1}{2}\nabla_{\al}Y\nabla^{\al}Y}
{D8}
where $R$ again is the scalar curvature of $M_2$. We can put 
this result into the 4D Einstein-Hilbert action and then integrate over the 
isometric coordinates while decomposing the measure as 
\nobreak{$\sqrt{-g_{M_4}}=X\sqrt{-g_{M_2}}$}. The effective 2D action
divided by the (finite) volume of $T^2$ reads
\eq{
S=\int_{M_2}d^2x\sqrt{-g}\left[XR+\frac{(\nabla X)^2}{2X}
-\frac{1}{2}X(\nabla Y)^2\right] .}
{D11}
It is interesting
to note that the dilaton $X$ acquires the scaling factor of the 4D metric 
while the dilaton $Y$ represents the gravitational degree of freedom of 
the Gowdy model. Clearly this action is of the
Einstein form (\ref{3}). It can be shown that its variation leads to the 
same EOM as the ones from the original 4D action when the symmetry is 
introduced there. This point is nontrivial as witnessed 
by the reduced action resulting from warped metrics in Einstein gravity 
\cite{kkk98}.

From the Gowdy line-element alone it is not clear why to define
as dilaton fields $X$ and $Y$ as in (\ref{D4}). In principle other gauge
choices are possible. But in this representation the $X$-field alone carries 
the scale factor and hence the geometric information (``radius'') of the
Gowdy spacetime, while the $Y$-field represents a propagating degree
of freedom (``graviton''). It is only in these variables (modulo trivial field
redefinitions not mixing the dilatons) that the factorizability property, as 
defined below, is manifest.

\section{Classification of TDT in 2D}

In this section we will introduce useful notions, with respect to which 
we will classify TDT. As a word of warning we would like to emphasize 
the following point: 
Our definitions are not invariant under arbitrary field redefinitions. 

However, for theories with only one dilaton coming from
the higher-dimensional metric field redefinitions, leading
to a mixing between the geometric dilaton and the ``scalar field'' dilaton 
in the new variables, are very inconvenient for two reasons:
First, in general such transformations change the geodesics of 
testparticles\footnote{The Christoffel symbols in the transformed geodesic 
equation $\ddot{x}^{\al}+\Gamma^{\al}{}_{\beta\gamma}(\tilde{X},\tilde{Y})
\dot{x}^{\beta}\dot{x}^{\gamma}=0$ become dependent on the new {\em matter} 
dilaton $\tilde{Y}$.} and hence the geometric properties of the manifold. 
Second, the quantization procedure used in \cite{klv99,gkv00} is
spoilt by a mixing of geometric and matter variables\footnote{In the following 
we will apply conformal transformations on such models. The conformal 
invariance (if any) then permits a ``mixing'' of the variables.}.

For the case where both dilatons stem from the higher-dimensional metric 
such field redefinitions correspond to the choice of a different gauge for 
this metric and are therefore possible. Using this feature, by choosing a 
specific ``gauge'', we will show below that a general 
class of such models always fits nicely into our classification scheme.

\subsection{Definitions}

We start the classification with some new definitions which prove useful:\\

{\bf Definition 1:} A TDT in the EF (\ref{3}) is called
{\em simple} iff $V_3^E = 0$.\\

Simple theories are models with no dynamical mixing between the dilaton 
fields and can be treated like a one-dilaton theory with dilaton field $X$
and an additional scalar matter field $Y$ coupled non-minimally in general.
A main reason for highlighting such models is the possibility to bring
them into a first order form (see below). This means that one can easily
apply the quantization scheme developed in this framework \cite{klv99}. 
Note that it is always possible to redefine the dilaton fields such 
that the diagonal term vanishes. E.g. one could use the nonlinear
transformation $X\to \tilde{X}^{\alpha}\cdot f(\tilde{\phi}),
\,\phi\to \tilde{X}^{1-\alpha}\cdot f^{-1}(\tilde{\phi});\, 
\alpha=(w+1)/(w+3/2)$ to make spherically reduced JBD ((\ref{D14}) 
for $w=const.$) simple. However, as we have already mentioned, such a 
redefinition is only allowed if both dilatons stem from the higher-dimensional
metric.\\

{\bf Definition 2:} A TDT in any given form is called {\em factorizable} iff 
$V_1 (X,Y) = f_1(X) g(Y)$ and $V_0 (X,Y) = f_2(X) g(Y)$. We assume that (at
least) $X$ is part of a higher-dimensional metric, while $Y$ can either be also
part of this metric or a ``true'' scalar field. \\

Factorizable theories permit a simple geometrical interpretation of $X$ as
``classical dilaton field'' in the 2D model, since there is a common 
$Y$-factor $g(Y)$ in front of the first two terms of (\ref{1}).
In the EF this property translates into $V_1^E = V_1^E(X)$. 

In the following we discuss a class of models (including all models
considered in this paper) where the original D-dimensional
spacetime $M_D$ contains one or two maximally symmetric, spacelike subspaces 
$S_{(1,2)} $(e.g. $S^1, S^2\otimes S^3$) such that locally 
$M_D\cong M_2\otimes S_{(1)}\otimes S_{(2)}$.

In the first case one dilaton arises from
the metric and the other from the matter Lagrangian. These models are 
trivially factorizable (e.g. JBD, SRG). 

In the second case both dilatons stem
from the metric which can then be written in the form \cite{wei72}
\footnote{As explained above this is just a specific choice of gauge 
for the metric of $M_D$.}
\begin{equation}
ds^2=g_{\alpha\beta}dx^{\alpha}dx^{\beta}-X^2[e^{2Y/d_1}d^2\Omega_{(1)}
+e^{-2Y/d_2}d^2\Omega_{(2)}]\label{fac1}
\end{equation}
where $d^2\Omega_{(1,2)}$ are the metrics of the two subspaces with
dimensions $d_1,d_2$ respectively, $g_{\alpha\beta}$ is the metric of the
reduced two-dimensional spacetime and $X(x^{\alpha}),Y(x^{\alpha})$ are the 
dilatons. The factors in the definition of $Y$ are chosen such that the 
measure only depends on $X$ which carries the conformal weight: 
\begin{equation}
\sqrt{(-1)^{D+1}g_{M_D}}=\sqrt{-g_{M_2}}\sqrt{(-1)^{d_1+d_2}
g_{(1)}g_{(2)}}\cdot X^{d_1+d_2}\,.\label{measure}
\end{equation}
The reduced two-dimensional action divided by the volumes of the subspaces 
reads
\begin{eqnarray}
S_2&=&\int_{M_2}\sqrt{-g_{M_2}}\cdot X^{d_1+d_2}
\left[R+(d_1+d_2)(d_1+d_2-1)
\frac{(\nabla X)^2}{X^2}-\right.\nonumber\\
&&\left.-(\frac{1}{d_1}+\frac{1}{d_2})(\nabla Y)^2-
\frac{R_{(1)}}{X^2}e^{-2Y/d_1}-\frac{R_{(2)}}{X^2}e^{2Y/d_2}\right]\label{fac2}
\end{eqnarray}
where a $\square X$ term has been partially integrated using the measure 
(\ref{measure}). $R_{(1,2)}$ are the (constant) scalar curvatures of the 
corresponding subspaces.
As the kinetic term of the $X$-dilaton does not depend on $Y$ we conclude
that also this class of models is factorizable, containing polarized Gowdy 
($T^2\cong S^1\otimes S^1$) and spherically reduced KKT ($S^1\otimes S^2$). 
From (\ref{fac2}) one can also see that these models are simple.
Thus, by field redefinitions -- corresponding to the choice of adapted
coordinates in the higher-dimensional metric -- it is always possible to
bring them into the EF and make the mixed kinetic term vanish.\\

From JBD we know that it is conformally equivalent to Einstein gravity
modulo the aforementioned problem of coupling to matter and a potential 
change of geodesic behavior. We will call such theories conformally related:\\

{\bf Definition 3:} Two theories are called {\em conformally related}, iff 
there exists a conformal transformation (in the higher-dimensional theory) 
between them.\\ 

It is an interesting task to investigate whether a TDT given in the JF is
conformally related to a simple model, since such models are 
particularly easy to treat and interpret. However, not all models allow a 
simplification through conformal transformations.\\ 

{\bf Definition 4:} If a non-simple model is conformally related to a simple 
model we call it {\em conformally simple}.\\

By explicit calculation we will show below that all factorizable, non-simple 
models where the functions $V_0,V_3$ are monomials are conformally simple, 
provided that only one of the dilatons carries a conformal weight.

We would like to emphasize that conformally related
theories represent dynamically inequivalent models in general. A simple
example is the CGHS-model \cite{cae92} which can either be
introduced by complete spherical reduction from an
infinite-dimensional Einstein-Hilbert action (cf.  eq.  (\ref{D3}) for 
$D \to \infty$) or by the requirement of scale-invariance in the 2D action:
\newline 
\parbox{3cm}{\begin{eqnarray*} V_0^{CGHS} &=& X, \\ V_3^{CGHS} &=& 0, 
\end{eqnarray*}} \hfill
\parbox{3cm}{\begin{eqnarray*} V_1^{CGHS} &=& 1/X, \\ V_4^{CGHS} &=& X, 
\end{eqnarray*}} \hfill
\parbox{3cm}{\begin{eqnarray*} V_2^{CGHS} &=& 0, \\  V_5^{CGHS} &=& 0. 
\end{eqnarray*}} \hfill
\parbox{1cm}{\begin{equation} \label{CGHS} \end{equation}} \hfill
\newline
This action is invariant under a constant rescaling $X \to \la X$. Through an 
intrinsically 2D conformal transformation with a conformal factor 
$\Om = X^{1/2}$ one can get rid of the $V_1$-term and the transformed theory 
describes flat spacetime.  Thus the conformally related global structures are 
profoundly different: The Black Hole singularity of CGHS has disappeared. 
Also for any other theory important properties of the spacetime such as the 
2D curvature and geodesic (in)completeness can be changed by a conformal 
transformation \cite{kst96b}.  

Despite of this, conformal transformations are frequently used 
in the literature on quantization of 2D dilaton gravity 
(cf.\  e.g.\  \cite{cau00}) or JBD (cf.\  e.g.\  
\cite{qui99}) although by now even some of the proponents of
this method \cite{cnt96} have (re)discovered this subtlety
\cite{cfn99}.  The issue of (in)equivalence of conformal
frames has a long history of confusion, as pointed out in
\cite{mas94, fgn98} (see also references
therein and references 28,29 of \cite{kuv99} for positive and 
negative examples).

It is necessary to bear in mind that most 
2D models are dimensionally reduced
theories which follow from a physically motivated higher-\-dimensional 
model. Another alternative is that they are 
merely toy models.  In both cases a conformal transformation changing
the global structure leads to a different theory. In the first 
case no longer  a 2D equivalent of the
original theory (the ``correct'' conformal frame is known) 
is described. In the second case,  one could have started from the
transformed toy model instead of introducing an
``auxiliary'' toy model (one could have introduced the
``correct'' conformal frame from the very beginning).
Of course, from a technical point of view, conformal transformations are very 
useful and indeed will be employed amply below, if they only 
represent an intermediate step (especially in the context of 
a classical theory: For the quantum case the frame where the 
quantization is performed must be the ``correct'' one under 
all circumstances). 

In the following we consider conformal transformations
\eq{
g_{\al\be} \to g_{\al\be}\Om^{-2}, \hspace{0.5cm}\sqrt{-g}\to \sqrt{-g}
\Om^{-2}, \hspace{0.5cm}X \to X\Om^{-\al} , \hspace{0.5cm}Y\to Y,
}{42}
assuming that $X$ has a conformal weight of $\al \in \mathbb{R}^*$ and $Y$ 
has conformal weight zero. This applies to models in the JF where 
only $X$ stems from dimensional reduction. To include a larger class of models
we generalize the JF by allowing $V_0$ to be an arbitrary function of $Y$:
\eq{
V_0 (X,Y) = X v_0(Y)\; . 
}{17}
Since we want to answer the question whether a model is conformally 
related to a simple one we have to impose the condition 
$V_3^E = 0$ after the (higher-dimensional) conformal transformation. 
The choice of the conformal factor 
\eq{
\Om = (v_0(Y))^{1/\al} 
}{5}
is necessary to bring the action in the EF.
The first term in (\ref{1}) as a consequence of the identity, 
valid under the conformal transformation (\ref{42}), 
\eq{
R \to R\Om^2 + 2 \frac{\nabla_{\ga} \nabla^{\ga} \Om}{\Om} - 2 
\frac{\nabla_{\ga} \Om \nabla^{\ga} \Om}{\Om^2}
}{8}
produces the first term in the EF (\ref{2}), plus further ``kinetic'' terms. 
Note that $\Om$ must be $C^1$, manifestly positive and invertible 
with respect to $Y$. Its inverse will be denoted by
\eq{
Y = f(\Om) .
}{5a}

\subsection{Conformally simple TDT}

In the following steps we will restrict ourselves to a smaller subset of TDT
having the advantage of simplifying calculations drastically while still
being general enough to include all ``physical'' models considered so far 
(and more). Using (\ref{8}) and dropping a boundary term the action (\ref{1}) 
after the conformal transformation (\ref{42}) with conformal factor (\ref{5}) 
becomes
\begin{multline}
S_c = \int_{M_2} d^2x \sqrt{-g} \\
\left[ X R - 2 \nabla_{\ga} \Om \nabla^{\ga} X + V_1 \Om^{-2\al} \left(\al^2 
X^2 \frac{(\nabla \Om)^2}{\Om^2} -2\al X  
\frac{\nabla_{\ga} X\nabla^{\ga} \Om}{\Om} + (\nabla X)^2 \right) \right. \\
\left. + V_2  {f'}^2 (\nabla \Om)^2 - \al V_3 X f' 
\Om^{-\al-1} (\nabla \Om)^2 + V_3 \Om^{-\al} f' 
\nabla_{\ga} X \nabla^{\ga} \Om  + V_4 \Om^{-2} \right] . \label{6}
\end{multline}
Note that in $V_i(X,Y)$ one has to replace $X \to X \Om^{-\al}$ and 
$Y \to f(\Om)$ as defined by (\ref{5a}).

Conformal simplicity requires the vanishing of the mixed term 
$\nabla_\ga Y \nabla^\ga X$. This yields a first order differential equation 
for the function $f$,
\eq{
f'(\Om) = \frac{2}{V_3} \left( \al V_1(X\Om^{-\al},\Om) X \Om^{-\al-1} + 
v_0(\Om) \Om^{-1}\right) ,
}{10}
which already restricts the functions $V_1,V_3$ severely because the l.h.s. of
(\ref{10}) is $X$-independent by construction. 

The convenient ansatz, to be used as of now,
\eq{
V_1 = X^{-1} v_1 (\Om), \hspace{0.5cm} V_3 = v_3 (\Om)
}{12}
is sufficient to satisfy (\ref{10}) although not necessary.
Next we impose factorizability on the original model which together with 
(\ref{12}) implies
\eq{
V_1 (X,\Om) = v_1 X^{-1} \Om^{\al}, \hspace{0.5cm} v_1 \in \mathbb{R} . 
}{18}
Assuming monomiality for $V_3$ by 
\eq{
v_3(\Om) = v_3 \Om^{\be}, \hspace{0.5cm} \be \in \mathbb{R}, \hspace{0.5cm}
v_3 \in \mathbb{R}^*
}{16}
the differential equation (\ref{10}) establishes a four-parameter 
solution\footnote{The solution for the conformal factor of course only 
depends on two real parameters, $c$ and $\ga$, but the original action and 
the transformed one contain all four parameters $\al$, $\be$, $v_1$ and $v_3$.}
\eq{
Y = f(\Om) = 2 \Om^{\al - \be} \frac{1 + \al v_1}
{v_3 (\al-\be)} =: c \Om^{\ga}, \hspace{0.5cm} \al \neq \be .
}{19}
This equation implies monomiality of $v_0(Y)$, too:
\eq{
v_0(Y) = c^{-\de} Y^{\de} , \hspace{0.5cm} \de := \frac{\al}{\al-\be} . 
}{19b}
Thus, with the assumptions made above $v_0(Y)$ is completely determined.

The resulting action may be written as
\eqa{
S_{cs} &=& \int_{M_2} d^2x \sqrt{-g} \left[ X R + v_1 \frac{(\nabla X)^2}{X} +
V_4\left(\frac{X}{v_0(Y)}, Y\right) v_0(Y)^{-2/\al} \right. \nonumber \\
&&\left. + \left(\nabla Y\right)^2 \left[ V_2 (\frac{X}{v_0(Y)}, Y) -
\al\de^2 \frac{X}{Y^2} (2+v_1\al) \right] \right] ,
}{6a}
which is conformally related to the original TDT action
\eqa{
S_{TDT} &=& \int_{M_2} d^2x \sqrt{-g} \left[ X v_0(Y) R + v_0(Y) v_1 
\frac{(\nabla X)^2}{X} + V_4\left(X, Y\right) \right. \nonumber \\
&&\left. + V_2 (X, Y) \left(\nabla Y\right)^2 + v_3 \frac{c v_0(Y)}{Y} 
\nabla_{\ga} X \nabla^{\ga} Y \right].}{6b} 
We recall the meaning of the four parameters: $\al$ is the conformal weight of
the (geometric) dilaton $X$, $\de$ (or $\be$ or $\ga$) defines the power of 
the monomial $v_0(Y)$, $v_1$ and $v_3$ are constants entering the 
corresponding functions ($c$ is defined in (\ref{19}) and depends on all these
constants).

Thus we have shown that a TDT satisfying (\ref{42})-(\ref{19b}) is conformally
simple, provided that $Y$ is positive everywhere. We emphasize that 
factorizability is conserved under this conformal transformation, as can be 
seen from (\ref{6a}), and hence is an independent property indeed.

The most important examples are STT, which are known to be
conformally simple \cite{jor55, fie56, brd61}:
\eq{
\al = 2,\, \de = 1 (\be = 0),\, v_1 = 1/2,\, v_3 = 2\,.
}{20}
The relation between the conformal factor and the STT field is 
$f(\Om) = Y = \Om^2$, a well-known result. Note that the whole class of 
STT is given by a single point in the four-dimensional parameter space 
of possible conformally simple actions with conformal factor given by 
(\ref{19}). Thus, despite of the various restrictions which led to 
(\ref{19}), the set of conformally simple theories described by the action 
(\ref{6a}), resp. (\ref{6b}), is very large. 

It is straightforward to construct toy models
which possess all possible combinations of factorizability and (conformal)
simplicity.

\section{First Order Formalism}

Dilaton  models that are (conformally) simple as well as factorizable 
have the important property that they may be written in an 
equivalent first order form \cite{iki93}. 
This is manifest in the EF.  In this case one of the dilatons (namely the 
scalar field $Y$) is disentangled from the gravitational sector, in the sense 
that no mixed kinetic term appears. The geometric part of the Lagrangian 
(including the dilaton $X$) can be brought to first order in its derivatives 
by introducing Cartan variables and auxiliary fields $X^a$. The zweibein basis 
is expressed in light-cone coordinates\footnote{We choose a representation 
$(0,1)\rightarrow(-,+)$. Light-cone indices are adorned with bars.} 
$e^{\mp}=\frac{1}{\sqrt{2}}\left(e^0\mp e^1\right)$.
The invariant volume element in this frame is given by 
$d^2x\sqrt{-g}=d^2x(e)=e^-\wedge e^+$. The Levi-Civit{\'a} symbol 
$\eps^{\bar{a}\bar{b}}$ is defined by\footnote{The 
$\eps$-symbol in ordinary coordinates is defined by $\eps^{01}=1$.} 
$\eps^{-+}=-1$.  The connection 1-form $\omega^{\bar{a}}_{\,\,\bar{b}}$ 
which is proportional to $\eps^{\bar{a}}_{\,\,\bar{b}}$ becomes
$\omega^{\bar{a}}_{\,\,\bar{b}}= -\om \cdot \text{diag}\left(1,-1\right)$.
Thus the 2D scalar curvature can be written as
$d^2x\sqrt{-g}R=-2d\omega$.  According to the second reference of \cite{iki93} 
we add the terms $X_{\bar{a}}T^{\bar{a}}=X^-(d+\omega)e^++X^+(d-\omega)e^-$
to the action where $T^{\bar{a}}$ is the torsion associated
with the connection $\omega$. The EF action (\ref{2}) divided by
$(-2)$ becomes equivalent to
\begin{multline}
\hspace{-0.4cm}S_{FO} = \underset{M_2}{\int}\left[X^-\left(d+\omega\right)e^+
+X^+\left(d-\omega\right)e^-+Xd\omega +e^-\wedge e^+V^E_1\left(X\right)X^-X^+-
\right. \\
\hspace{0.3cm}\left.-\frac{1}{2}V^E_2\left(X,Y\right)dY\wedge\ast dY-
\frac{1}{2}e^-
\wedge e^+\left(V^E_4\left(X,Y\right)+V^E_5\left(X,Y\right)f_m\right)\right] ,
\label{P6}\end{multline}
where we have included also the matter term. The fields
$X^{\mp}$ and $X$ are determined from the EOM produced by the 
variation of the Cartan variables. 
The whole set of EOM derived from (\ref{P6}) is equivalent
to the one obtained from the original action
\cite{klv97a}. Actually $X^{\pm}$ and $\omega$ may be 
eliminated by {\em algebraic} EOM from (\ref{P6}). 

For a theory in the EF the corresponding first order formulation has 
many advantages, especially at the quantum level, where 
e.g.\ the geometric degrees of freedom of SRG can be 
integrated exactly \cite{klv97a,klv99}.  Here we will
only use one important result, namely the existence of a
conservation law that can be derived in a particularly
simple way in this context \cite{kus92,kut98, grk00}.
Taking appropriate linear combinations of the EOM derived from (\ref{P6})
with an integrating factor $I(X) = e^{-\int^XV^E_1\left(X^{\prime}\right)
dX^{\prime}}$ we obtain a relation of the type
\eq{
I(X)\partial_{\alpha}\left(X^-X^+\right)-I(X)\partial_{\alpha}X\left(V^E_1
\left(X\right)X^-X^+-\frac{V^E_4\left(X,Y\right)}{2}\right) + W_{\al}=0.
}{P7}
Splitting the potential $V^E_4$ into two terms 
$V^E_4=V^{\left(g\right)}_4\left(X\right)+V^{\left(Y\right)}_4\left(X,Y\right)$
we obtain the conservation law
\eq{
\partial_{\al}{\cal C}=\partial_{\al}{\cal C}^{\left(g\right)}+W_{\al} = 0 
}{P8}
where ${\cal C}^{\left(g\right)}=X^-X^+I(X)+\frac{1}{2}\int^X
V^{\left(g\right)}_4\left(X^{\prime}\right) I(X^{\prime}) dX^{\prime}$. 
From (\ref{P8}) the 1-form $W_{\al} = W^{(Y)}_{\al} + W^{(m)}_{\al}$ is 
trivially exact.
Its separation into matter and $Y$-terms depends on the coupling function 
$V_5^E$ that can have an arbitrary $Y$-dependence. The components 
$W^{\left(Y\right)}_{\al}$ are given by
\meq{
W^{\left(Y\right)}_{\al} = I(X) \left[\frac{\partial_{\alpha}X}{2}
V^{\left(Y\right)}_4\left(X,Y\right)+\right. \\
+\left.\frac{V^E_2\left(X,Y\right)}{\left(e\right)^2}\left\{Y^-Y^+
\left(\partial_{\alpha}X\right)-\left(e\right)\left(Y^-X^++Y^+X^-\right)
\partial_{\alpha}Y\right\}\right]
}{P10}
where $Y^{\mp}=\eps^{\alpha\beta}e^{\mp}_{\,\,\beta}\left(\partial_{\alpha}
Y\right)$. The analogous expression for the matter part becomes
\eq{
W^{\left(m\right)}_{\al} = I(X) \frac{V^E_5\left(X,Y\right)}{2}
\left[\left(\partial_{\alpha}X\right)f_m-\left(e\right)\eps_{\alpha\beta}
\left(X^-\frac{\partial f_m}{\partial e^-_{\,\,\beta}}+X^+
\frac{\partial f_m}{\partial e^+_{\,\,\beta}}\right)\right] .}
{P11}
It has been shown \cite{grk00} that this conservation law is connected to the 
energy conservation of the model considered. More precisely, the geometric 
part ${\cal C}^{(g)}$ is proportional to a mass-aspect function $m_{eff}(r,t)$ 
which is the sum of the ADM mass and the energy fluxes given by the matter- 
and $Y$-contributions. Since we have not specified as yet the functions 
$V^E_1,V^E_2,V^E_4,V^E_5$ we have generalized that conservation law from 
EMKG to all factorizable (conformally) simple theories.

It is interesting to check its behavior under a (constant) Weyl-rescaling 
$g_{\alpha\beta}\rightarrow\lambda g_{\alpha\beta}$ in the 
higher-\-dimensional theory (taking into account the conformal weight of the 
geometric dilaton $X$). For general 
spherically reduced models we obtain a scaling weight
\eq{
\left. \text{sw}({\cal C}) \right|_{SR} = D-3 ,}
{cw}
where D is the higher dimension (e.g. D=4 for ordinary SRG), provided 
that the matter part and $Y$-part couple linearly to the dilaton $X$. In all
other cases (the coupling to matter or $Y$ is different or the potentials
differ from SRG) the conserved quantity does not have a well defined
conformal weight with respect to a global conformal transformation in the 
higher-\-dimensional theory (e.g. the CGHS model (\ref{CGHS}) contains 
minimally coupled scalar fields instead of linearly coupled ones).

\section{Conclusions}

In this paper we have investigated TDT produced by dimensional reduction.
First we introduced two standard forms, namely the Einstein form (\ref{2})
and the Jordan form (\ref{3}), covering all models considered in this paper.
The useful properties factorizability, simplicity and conformal 
simplicity have been defined. Since there seems to be still confusion 
in the literature (for a selected list of such papers cf. e.g. the review 
article \cite{fgn98}) we have emphasized the physical inequivalence of 
conformally related theories. 

We could show that all investigated models can be derived by dimensional 
reduction of a spacetime with one or two maximally symmetric 
subspaces\footnote{In some cases like JBD one ``forgets'' about the origin of 
one of the dilatons. This manifests itself in setting its conformal weight 
equal to zero.}. From this we conclude that:
\begin{itemize}
\item All these models are factorizable.
\item Models with one geometric dilaton are at least conformally simple.
\item Models with two geometric dilatons are simple (in adapted coordinates).
\end{itemize}
The second result was obtained by explicitly mapping a non-simple TDT
in the JF (with $V_0,V_3$ monomials) onto a simple TDT in the EF through a 
conformal transformation.
Thereby we could also show that factorizability is conserved under such
a conformal transformation.
It is not yet clear if theories with more complicated Killing-orbits
are still factorizable. This could be the subject of further work.

\begin{sidewaystable}
\centering
\fbox{
\begin{tabular}{|l||c|c|c|c|c|c||c|c|c|}\hline
Model       & Form & $V_1$    & $V_2$      & $V_3$ & $V_4$    & $V_5$  
& Simple & DOF \\ \hline \hline
1 dilaton   & EF & $V_1(X)$ & $0$        & $0$   & $V_4(X)$  & $0$     
& Yes   & $0$ \\ \hline
Bos. String & EF & $V_1(X)$ & $const.$   & $0$   & $V_4(X)$  & $1$     
& Yes & $N$\\ \hline
SR EMKG     & EF & $1/2X$   & $-\ka X$   & $0$   & $-1/2$    & $0$     
& Yes   & $1$ \\ \hline
Pol. Gowdy  & EF & $1/2X$   & $-1/2X$    & $0$   & $0$       & $0$     
& Yes & $1$ \\ \hline
SR $R^2$    & JF & $Y/2X$   & $0$        & $2$   & $-Y/2-XY^2/2$ & $0$ 
& Conf. & $0$ \\ \hline
SR KKT      & EF & $2/3X$   & $-3X/2$      & $0$ & $-2\sqrt[3]{X}e^{-Y}$    
& $-X$ & Yes  & $1+m$ \\ \hline
SR CCS      & JF & $Y/2X$   & $(3/2)X/Y$    & $2$   & $-Y/2$    & $0$     
& Conf. & $1$ \\ \hline
SR JBD      & JF & $Y/2X$   & $-w X/Y$ & $2$   & $-Y/2$    & $-X$    
& Conf. & $1+m$ \\ \hline
SR STT(J)   & JF & $Y/2X$ & $-w X/Y$ & $2$ & $-Y/2+X\hat{V}(Y)$ & $-X$ 
& Conf. & $1+m$ \\ \hline
SR STT(J) & EF & $1/2X$ & $-(w+3/2)X/Y^2$ & $0$ & $-1/2+X\hat{V}(Y)/Y^2$ & 
$-X/Y^2$ & Yes & $1+m$ \\ \hline  
SR STT(E) & EF & $1/2X$ & $-(w+3/2)X/Y^2$ & $0$ & $-1/2-X\hat{V}(Y)$ & $-X$ 
& Yes & $1+m$ \\ \hline
SR STT(E) & JF & $Y/2X$ & $-w X/Y$ & $2$ & $-Y/2+X\hat{V}(Y)Y^2$ & $-XY^2$ 
& Conf. & $1+m$ \\ \hline
\end{tabular}} 
\caption[table1]{A representative sample of TDT and their properties. 
Abbreviations: SR stands for spherically reduced. 
DOF means (continuous physical) degrees of freedom ($m$ denotes 
the number of matter DOF); N is the number of target space coordinates for the 
bosonic string. In the case of 1 dilaton models or the bosonic 
string by adjusting $V_1(X)$ and $V_4(X)$ one obtains a variety 
of models, among them the CGHS model \cite{cae92},
the Jackiw-Teitelboim model \cite{tei83} and the 
Katanaev-Volovich model \cite{kvo86}. CCS means conformally coupled
scalar in D=4. This model is the limit $w \to -3/2$ of JBD. In 
the entries of spherically reduced STT(X) minimal coupling to matter in the 
X-frame in D=4 has been assumed. All models are factorizable.}
\label{tab:table1}
\end{sidewaystable}

Our subsequent investigations were restricted to the subclass of 
(conformally) simple factorizable theories. In the EF these models allow a 
first order formulation, and by straightforward application of previous work
\cite{kus92, man93, kut98, grk00} an absolute conservation law (\ref{P8}) 
could be established. 

By investigating the scaling weight of the conserved quantity under global 
conformal transformations in the higher-dimensional theory for SRG the 
intuitively expected result (\ref{cw}) was obtained. Moreover it was clarified
that a necessary condition for a definite scaling weight of the conserved 
quantity was linear coupling of the matter fields and the second dilaton $Y$ 
to the (geometric) dilaton $X$, i.e. to the dilaton with a non-vanishing 
scaling weight.

Apart from the obvious applications (namely a 2D description of various
higher-dimensional models considered in this paper) TDT serve as toy models 
for D=4 theories with two dilaton fields and as a basis for generalizations
to models with more than two dilatons. Compactification of e.g. 
D=11 supergravity can yield two or more dilaton fields, and up to now no 
satisfactory cosmological theory with a single scalar field (which serves e.g. 
as inflaton {\em and} quintessence) is known. Here one may hope
that -- as in the case of the nonvanishing cosmological constant (or
quintessence?) -- further input may be provided by the enormously 
increasing amount of astrophysical data to be expected for the near future.
If the need for more dilaton fields should arise we believe that similar 
structures in the classification of such models will appear.

Although the physically relevant TDT are related to dimensional reduction it
could be of interest to focus on an intrinsically 2D treatment of TDT. Such an 
investigation would e.g. involve a conformal transformation where no conformal 
weight is attributed to both dilatons and suggest the introduction of a 
``true'' Jordan frame, i.e. a conformal frame where no dilaton at all is 
coupled to the scalar curvature.

At the quantum level the next step should be a Hamiltonian analysis 
and BRST quantization. Similarities to the analysis of non-minimally 
coupled scalars interacting with a one-dilaton theory \cite{gkv00a} 
which is based upon the simpler results obtained for the minimally 
coupled case \cite{klv97, klv99, gkv00} may well occur. In fact, for simple
factorizable theories the constraint algebra is already known \cite{gkv00a}
and differs only slightly from the simpler algebra obtained in \cite{klv97}.
Non-simple, but conformally simple factorizable models fit into this
theoretical frame only through a conformal transformation. Thus it will be an 
interesting task to investigate the action of a conformal transformation on 
the constraint algebra. This would provide a basis of (path integral) 
quantization of {\em all} conformally simple factorizable TDT.

\section*{Acknowledgements}

The authors have benefitted from discussions with D. Schwarz. One of the 
authors (D.G.) would like to thank P. H\"ubner for drawing his attention to 
Gowdy models. We are also grateful to the referee for his detailed criticism
which led to an improved manuscript.\\
This research has been supported by Project P14650-TPH of the Austrian Science
Foundation (\"Osterreichischer Fonds zur F\"orderung der 
wis\-sen\-schaft\-lichen Forschung).

\providecommand{\href}[2]{#2}\begingroup\raggedright
\endgroup


\begin{thebibliography}{10}

\bibitem{kal21}
T.~Kaluza, ``On the problem of unity in physics,'' {\em Sitzungsber. Preuss.
  Akad. Wiss. Berlin. Math. Phys.} {\bf K 1} (1921) 966;
O.~Klein, ``Quantum theory and five-dimensional relativity,'' {\em Z. Phys.}
  {\bf 37} (1926) 895; 
``The atomicity of electricity as a quantum theory law,'' {\em
  Nature} {\bf 118} (1926) 516.

\bibitem{fie56}
M.~Fierz, ``{\"U}ber die physikalische {D}eutung der erweiterten
  {G}ravitationstheorie {P.} {J}ordans,'' {\em Helv. Phys. Acta} {\bf 29}
  (1956) 128.

\bibitem{jor55}
P.~Jordan, {\em {S}chwerkraft und {W}eltall : {G}rundlagen der theoretischen
  {K}osmologie}.
\newblock Vieweg, second~ed., 1955;
``The present state of {D}irac's cosmological hypothesis,'' {\em Z.
  Phys.} {\bf 157} (1959) 112--121.

\bibitem{brd61}
C.~Brans and R.~H. Dicke, ``Mach's principle and a relativistic theory of
  gravitation,'' {\em Phys. Rev.} {\bf 124} (1961) 925--935;
R.~H. Dicke, ``Mach's principle and invariance under transformation of units,''
  {\em Phys. Rev.} {\bf 125} (1962) 2163--2167.

\bibitem{pal97}
{\bf The Supernova Cosmology Project} Collaboration, S.~Perlmutter {\em et
  al.}, ``Cosmology from type ia supernovae,'' {\em Bull. Am. Astron. Soc.}
  {\bf 29} (1997) 1351, \href{http://www.arXiv.org/abs/astro-ph/9812473}{{\tt
  astro-ph/9812473}};
``Measurements of omega and lambda from 42 high-redshift supernovae,''
  \href{http://www.arXiv.org/abs/astro-ph/9812133}{{\tt astro-ph/9812133}}.

\bibitem{zws99}
I.~Zlatev, L.~Wang, and P.~J. Steinhardt, ``Quintessence, cosmic coincidence,
  and the cosmological constant,'' {\em Phys. Rev. Lett.} {\bf 82} (1999) 896,
  \href{http://www.arXiv.org/abs/astro-ph/9807002}{{\tt astro-ph/9807002}}.

\bibitem{tih84}
P.~Thomi, B.~Isaak, and P.~H\'{a}j\'{\i}\v{c}ek, ``Spherically symmetric
  systems of fields and black holes. 1. definition and properties of apparent
  horizon,'' {\em Phys. Rev.} {\bf D30} (1984) 1168;
P.~H\'{a}j\'{\i}\v{c}ek, ``Spherically symmetric systems of fields and black
  holes. 2. apparent horizon in canonical formalism,'' {\em Phys. Rev.} {\bf
  D30} (1984) 1178;
S.~R. Lau, ``On the canonical reduction of spherically symmetric gravity,''
  {\em Class. Quant. Grav.} {\bf 13} (1996) 1541--1570,
  \href{http://www.arXiv.org/abs/gr-qc/9508028}{{\tt gr-qc/9508028}}.

\bibitem{iki93}
N.~Ikeda and K.~I. Izawa, ``Quantum gravity with dynamical torsion in
  two-dimensions,'' {\em Prog. Theor. Phys.} {\bf 89} (1993) 223--230;
P.~Schaller and T.~Strobl, ``Poisson structure induced (topological) field
  theories,'' {\em Mod. Phys. Lett.} {\bf A9} (1994) 3129--3136;
W.~Kummer and P.~Widerin, ``Conserved quasilocal quantities and general
  covariant theories in two-dimensions,'' {\em Phys. Rev.} {\bf D52} (1995)
  6965--6975, \href{http://www.arXiv.org/abs/gr-qc/9502031}{{\tt
  gr-qc/9502031}};
T.~Kl{\"o}sch and T.~Strobl, ``Classical and quantum gravity in
  (1+1)-dimensions, {P}art 1: {A} unifying approach,'' {\em Class. Quantum
  Grav.} {\bf 13} (1996) 965--984,
  \href{http://www.arXiv.org/abs/gr-qc/9508020}{{\tt gr-qc/9508020}}. Erratum
  ibid. 14 (1997) 825.

\bibitem{kus92}
W.~Kummer and D.~J. Schwarz, ``Renormalization of {$R^2$} gravity with
  dynamical torsion in {$d = 2$},'' {\em Nucl. Phys.} {\bf B382} (1992)
  171--186;
H.~Grosse, W.~Kummer, P.~Presnajder, and D.~J. Schwarz, ``Novel symmetry of
  noneinsteinian gravity in two- dimensions,'' {\em J. Math. Phys.} {\bf 33}
  (1992) 3892--3900, \href{http://www.arXiv.org/abs/hep-th/9205071}{{\tt
  hep-th/9205071}};
W.~Kummer and P.~Widerin, ``Non{E}insteinian gravity in d=2: {S}ymmetry and
  current algebra,'' {\em Mod. Phys. Lett.} {\bf A9} (1994) 1407--1414.

\bibitem{kut98}
W.~Kummer and G.~Tieber, ``Universal conservation law and modified {N}oether
  symmetry in 2d models of gravity with matter,'' {\em Phys. Rev.} {\bf D59}
  (1999) 044001, \href{http://www.arXiv.org/abs/hep-th/9807122}{{\tt
  hep-th/9807122}}.

\bibitem{grk00}
D.~Grumiller and W.~Kummer, ``Absolute conservation law for black holes,'' {\em
  Phys. Rev.} {\bf D61} (2000) 064006,
  \href{http://www.arXiv.org/abs/gr-qc/9902074}{{\tt gr-qc/9902074}}.

\bibitem{gow71}
R.~H. Gowdy, ``Gravitational waves in closed universes,'' {\em Phys. Rev.
  Lett.} {\bf 27} (1971) 826--829.

\bibitem{mas94}
G.~Magnano and L.~M. Sokolowski, ``On physical equivalence between nonlinear
  gravity theories and a general relativistic selfgravitating scalar field,''
  {\em Phys. Rev.} {\bf D50} (1994) 5039--5059,
  \href{http://www.arXiv.org/abs/gr-qc/9312008}{{\tt gr-qc/9312008}}.

\bibitem{dnp86}
M.~J. Duff, B.~E.~W. Nilsson, and C.~N. Pope, ``Kaluza-klein supergravity,''
  {\em Phys. Rept.} {\bf 130} (1986) 1.

\bibitem{dan93}
T.~Damour and K.~Nordtvedt, ``General relativity as a cosmological attractor of
  tensor scalar theories,'' {\em Phys. Rev. Lett.} {\bf 70} (1993) 2217--2219;
``Tensor - scalar cosmological models and their
  relaxation toward general relativity,'' {\em Phys. Rev.} {\bf D48} (1993)
  3436--3450.

\bibitem{kkk98}
M.~O. Katanaev, T.~Kl{\"o}sch, and W.~Kummer, ``Global properties of warped
  solutions in general relativity,'' {\em Annals Phys.} {\bf 276} (1999) 191,
  \href{http://www.arXiv.org/abs/gr-qc/9807079}{{\tt gr-qc/9807079}}.

\bibitem{klv99}
W.~Kummer, H.~Liebl, and D.~V. Vassilevich, ``Integrating geometry in general
  2{D} dilaton gravity with matter,'' {\em Nucl. Phys.} {\bf B544} (1999) 403,
  \href{http://www.arXiv.org/abs/hep-th/9809168}{{\tt hep-th/9809168}}.

\bibitem{wei72}
S.~Weinberg, ``GRAVITATION AND COSMOLOGY'', John Wiley and Sons, New York, 1972

\bibitem{cae92}
C.~G. {Callan Jr.}, S.~B. Giddings, J.~A. Harvey, and A.~Strominger,
  ``Evanescent black holes,'' {\em Phys. Rev.} {\bf D45} (1992) 1005--1009.

\bibitem{kst96b}
T.~Kl{\"o}sch and T.~Strobl, ``Classical and quantum gravity in
  (1+1)-dimensions, {P}art 2: {T}he universal coverings,'' {\em Class. Quantum
  Grav.} {\bf 13} (1996) 2395--2422,
  \href{http://www.arXiv.org/abs/gr-qc/9511081}{{\tt gr-qc/9511081}};
M.~O. Katanaev, W.~Kummer, and H.~Liebl, ``Geometric interpretation and
  classification of global solutions in generalized dilaton gravity,'' {\em
  Phys. Rev.} {\bf D53} (1996) 5609--5618;
``On the completeness of the black
  hole singularity in 2-{D} dilaton theories,'' {\em Nucl. Phys.} {\bf B486}
  (1997) 353--370.

\bibitem{cau00}
M.~Cavagli\`a, ``Quantum gravity corrections to the {S}chwarzschild mass,''
  {\em Phys. Rev.} {\bf D61} (2000) 064019,
  \href{http://www.arXiv.org/abs/hep-th/9912024}{{\tt hep-th/9912024}}.

\bibitem{qui99}
I.~Quiros, ``Is the {S}chwarzschild black hole spurious?,''
  \href{http://www.arXiv.org/abs/gr-qc/9903041}{{\tt gr-qc/9903041}}.

\bibitem{cnt96}
J.~Cruz, J.~Navarro-Salas, M.~Navarro, and C.~F. Talavera, ``Symmetries and
  black holes in 2{D} dilaton gravity,''
  \href{http://www.arXiv.org/abs/hep-th/9606097}{{\tt hep-th/9606097}}.

\bibitem{cfn99}
J.~Cruz, A.~Fabbri, and J.~Navarro-Salas, ``Can conformal transformations
  change the fate of 2{D} black holes?,'' {\em Phys. Lett.} {\bf B449} (1999)
  30, \href{http://www.arXiv.org/abs/hep-th/9811246}{{\tt hep-th/9811246}}.

\bibitem{fgn98}
V.~Faraoni, E.~Gunzig, and P.~Nardone, ``Conformal transformations in classical
  gravitational theories and in cosmology,''
  \href{http://www.arXiv.org/abs/gr-qc/9811047}{{\tt gr-qc/9811047}}.

\bibitem{kuv99}
W.~Kummer and D.~V. Vassilevich, ``Hawking radiation from dilaton gravity in
  (1+1) dimensions: {A} pedagogical review,'' {\em Annalen Phys.} {\bf 8}
  (1999) 801--827, \href{http://www.arXiv.org/abs/gr-qc/9907041}{{\tt
  gr-qc/9907041}}.

\bibitem{klv97a}
W.~Kummer, H.~Liebl, and D.~V. Vassilevich, ``Exact path integral quantization
  of generic 2-{D} dilaton gravity,'' {\em Nucl. Phys.} {\bf B493} (1997)
  491--502, \href{http://www.arXiv.org/abs/gr-qc/9612012}{{\tt gr-qc/9612012}}.

\bibitem{tei83}
C.~Teitelboim, ``Gravitation and hamiltonian structure in two space-time
  dimensions,'' {\em Phys. Lett.} {\bf B126} (1983) 41;
E.~D'Hoker, D.~Freedman, and R.~Jackiw, ``{SO(2,1)} invariant quantization of
  the liouville theory,'' {\em Phys. Rev.} {\bf D28} (1983) 2583;
R.~Jackiw, ``Lower dimensional gravity,'' {\em Nucl. Phys.} {\bf B252} (1985)
  343--356.

\bibitem{kvo86}
M.~O. Katanaev and I.~V. Volovich, ``String model with dynamical geometry and
  torsion,'' {\em Phys. Lett.} {\bf B175} (1986) 413--416.

\bibitem{man93}
R.~B. Mann, ``Conservation laws and 2-{D} black holes in dilaton gravity,''
  {\em Phys. Rev.} {\bf D47} (1993) 4438--4442,
  \href{http://www.arXiv.org/abs/hep-th/9206044}{{\tt hep-th/9206044}}.

\bibitem{gkv00a}
D.~Grumiller, W.~Kummer, and D.~Vassilevich, ``Perturbation theory for
  non-minimally coupled scalar matter (preliminary title).'' (in preparation).

\bibitem{klv97}
W.~Kummer, H.~Liebl, and D.~V. Vassilevich, ``Nonperturbative path integral of
  2d dilaton gravity and two loop effects from scalar matter,'' {\em Nucl.
  Phys.} {\bf B513} (1998) 723--734.

\bibitem{gkv00}
D.~Grumiller, W.~Kummer, and D.~Vassilevich, ``The virtual black hole in 2d
  quantum gravity.'' {\em {N}ucl. {P}hys.} {\bf B580} (2000) 438--456,
 \href{http://www.arXiv.org/abs/hep-th/0001038}{{\tt hep-th/0001038}}.

\end{thebibliography}
\end{document}